\documentclass[fp]{jpsj3}

\usepackage{txfonts}

\title{Adiabatic Approximation for the Imaginary-Time Schr\"odinger Equation
and Its Application to Simulated Annealing}
\author{Kazuya Kaneko$^1$
and Hidetoshi Nishimori$^2$}
\inst{$^1$Department of Basic Science, The University of Tokyo, Meguro, Tokyo, 153-8902, Japan\\
$^2$Department of Physics, Tokyo Institute of Technology, Meguro, Tokyo 152-8551, Japan}

\abst{
We formulate an adiabatic approximation for the imaginary-time Schr\"odinger equation.
The obtained adiabatic condition consists of two inequalities,
one of which coincides with the conventional adiabatic condition for the real-time Schr\"odinger equation,
but the other does not.
We apply this adiabatic approximation to
the analysis of Markovian dynamics of the classical Ising model, which can be formulated as the
imaginary-time Schr\"odinger equation, to obtain an asymptotic formula for the probability that
the system reaches the ground state in the limit of a long annealing time in simulated annealing.
Using this form, we amend the theory of Somma, Batista, and Ortiz
for a convergence condition for simulated annealing.
}

\begin{document}

\maketitle

\section{Introduction}
\label{sec:Intro}
An optimization problem is a problem of finding an element of some set
that minimizes a real-valued function called the cost function.
In this paper, we consider an optimization problem with discrete variables,
which is known as a combinatorial optimization problem.
The cost function of a combinatorial optimization problem is identified with the Hamiltonian of the classical Ising model
whose ground state is the global minimum.
Solving combinatorial optimization problems is difficult in general
because of the exponential increase of the number of elements with the problem size and frustrations in the problem.
It is generally very difficult to find the exact solution within a practical time.
We thus devise algorithms that give an approximate solution.
Simulated annealing~\cite{SA,Spinglass} and quantum annealing~\cite{QA,Santoro,Das} are among such approximate algorithms.

The basic idea of these algorithms is to use a physical process
to escape local minima of the cost function so that the state approaches the global minimum.
In simulated annealing, we introduce a time-dependent temperature $ T(t) $ as the control parameter.
We initially set the temperature to a high value and reduce $ T(t) $  slowly toward zero,
and the system finally reaches the zero-temperature equilibrium,
the ground state that corresponds to the solution of the combinatorial optimization problem.
Quantum annealing was proposed in an analogy with simulated annealing~\cite{QA}. 
In quantum annealing, we  introduce a time-dependent external magnetic field
which induces quantum fluctuations.
We reduce the external magnetic field from a very large value to zero,
similar to simulated annealing in which we reduce the temperature.
A similar idea, adiabatic quantum computation~\cite{AQC}, is often used
in the literature of quantum information theory.
However, there is a small difference between adiabatic quantum computation and quantum annealing
in that adiabatic quantum computation only uses adiabatic time evolution,
but nonadiabatic time evolution is also considered in quantum annealing~\cite{Non-adi}.
In this paper, we consider quantum annealing following adiabatic time evolution,
\textit{i.e.}, adiabatic quantum computation.

The classical-to-quantum mapping discussed in Refs.~\citen{Henley,Castelnovo,Somma} allows us to express
the thermodynamical properties of classical systems in terms of those of quantum systems in the same spatial dimension.
Using this mapping, a slow change in the temperature in simulated annealing corresponds to
a slow change in  the Hamiltonian in quantum annealing.
Therefore, we can study simulated annealing  and quantum annealing from the same perspective. 
However, the mapped quantum state does not follow the real-time Schr\"odinger equation,
and its dynamics is represented as the imaginary-time Schr\"odinger equation~\cite{SA-QA}.
When we consider time-dependent quantities, we need careful analyses owing to the difference of the dynamics. 
Somma~\textit{et al.} applied this classical to quantum mapping to simulated annealing,
and rederived the convergence condition originally proved in Ref.~\citen{Geman}
under the ordinary adiabatic condition for the real-time Schr\"odinger equation~\cite{Somma}.
However, the real-time adiabatic condition does not directly apply
to the imaginary-time Schr\"odinger equation,
and their analysis should therefore be carefully reexamined.

In this work, we derive formulas for the adiabatic approximation for the imaginary-time Schr\"odinger equation.
This was derived before in Ref.~\citen{Morita} for the norm-conserved but nonlinear case as well as in Ref.~\citen{Grandi}.
Our approximation is applicable to the norm-nonconserved and linear case,
which is related to classical Markovian dynamics.
We apply this approximation to simulated annealing
and obtain an asymptotic formula for the probability that the system reaches the ground state at zero temperature.
Using this formula, we rederive the rate of convergence to the ground state discussed by Somma~\textit{et~al}.

This paper is organized as follows.
In the next section, we derive the adiabatic approximation for the imaginary-time Schr\"odinger equation.
Then, in Sect.~\ref{sec:CtoQ}, we review classical-to-quantum mapping
and rewrite classical Markovian dynamics as the imaginary-time Schr\"odinger equation.
Applying  the approximation discussed in Sect.~\ref{sec:IS} to the mapped quantum system,
we analyze the probability of reaching the ground state in Sect.~\ref{sec:prob}.
The convergence condition for simulated annealing is rederived
from the imaginary-time adiabatic condition  in Sect.~\ref{sec:cc}.
The final section is devoted to the conclusion.

\section{Imaginary-Time Schr\"odinger Equation and Its Adiabatic Approximation}
\label{sec:IS}

Let us consider the imaginary-time Schr\"odinger equation
\begin{equation}
-\frac{d}{dt}|\psi (t)\rangle =H(t)|\psi (t)\rangle .
\label{i-Schrodinger}
\end{equation}
We consider the time development of a system following this equation in
the time scale $\tau$, $0\le t\le \tau$.
We scale the time as $s=t/\tau$, where $s$ starts from 0 and ends at $s=1$.
Then, Eq.~(\ref{i-Schrodinger}) reads
\begin{equation}
-\frac{d}{ds}|\psi (s)\rangle =\tau H(s)|\psi (s)\rangle.
\label{i-Schrodinger2}
\end{equation}
Note that the norm of the wave function is not conserved, and
$\langle \psi (s)|\psi (s)\rangle$ depends on $s$.

Following Refs.~\citen{Morita} and \citen{Grandi},
we expand the wave function in terms of the set of instantaneous eigenstates,
$H(s)|j (s)\rangle =E_j(s)|j (s)\rangle$, as
\begin{equation}
|\psi (s)\rangle =\sum_j c_j(s)|j (s)\rangle =\sum_j e^{-\tau\phi_j (s)}\tilde{c}_j(s)|j (s)\rangle,
\end{equation}
where the second equality defines $\tilde{c}_j(s)$ with
\begin{equation}
\phi_j(s)=\int_0^s ds' E_j(s') .
\end{equation}
We assume $E_0(s)=0$ with an application in later sections in mind.
The imaginary-time Schr\"odinger equation~(\ref{i-Schrodinger2}) can be rewritten 
in terms of the coefficients as
\begin{equation}
\frac{d \tilde{c}_j(s)}{ds}=\sum_{k\ne j} e^{\tau (\phi_j(s)-\phi_k(s))} \frac{\langle
j(s)|\frac{dH(s)}{ds}|k(s)\rangle}{\Delta_{jk}(s)} \tilde{c}_k(s),
\end{equation}
where $ \Delta_{jk}(s)=E_j(s)-E_k(s) $.
Integration of this differential equation and multiplication of the resulting
expression by $e^{-\tau \phi_j(s)}$ yield
\begin{align}
c_j(s)&=c_j(0)e^{-\tau \phi_j(s)}+e^{-\tau \phi_j(s)}\sum_{k\ne j} \int_0^s ds'\,
e^{\tau \phi_j(s')}\frac{\langle j(s')|\frac{dH(s')}{ds'}|k(s')\rangle}{\Delta_{jk}(s')} c_k(s') .
\label{integralequation}
\end{align}

Let us solve this integral equation iteratively, \textit{i.e.}, an asymptotic
expansion for very large $\tau$.
The initial condition is that $c_0(0)$ for the ground state is of $\mathcal{O}(1)$,
and the other coefficients are much smaller or even zero.
Then, the zeroth-order solution $ c_j^{(0)} $,
which is obtained by ignoring the integral part in Eq.~(\ref{integralequation}), is
\begin{equation}
c_0^{(0)}(s)=c_0(0),\quad
c_{j(\ne 0)}^{(0)}=c_j(0)e^{-\tau \phi_j(s)}.
\end{equation}
Insertion of these relations into Eq.~(\ref{integralequation}) gives
\begin{align}
c_{j(\ne 0)}^{(1)}(s)
&=c_j(0)e^{-\tau \phi_j(s)}+e^{-\tau \phi_j(s)}
\sum_{k \ne j} c_k(0) \int_0^s ds'\, e^{\tau (\phi_j(s')-\phi_k(s'))}
\frac{\langle j(s')|\frac{dH(s')}{ds'}|k(s')\rangle}{\Delta_{jk}(s')}\\
&=c_0(0) e^{-\tau \phi_j(s)}\int_0^s ds' \, e^{\tau \phi_j(s')}
\frac{\langle j(s')|\frac{dH(s')}{ds'}|0(s')\rangle}{\Delta_{j0}(s')}+\mathcal{O}(e^{-\tau}).
\end{align}
Integration by parts leads to
\begin{align}
c_{j(\ne 0)}^{(1)}(s)&=c_0(0) e^{-\tau \phi_j(s)}\left\{\frac{1}{\tau}
\left[e^{\tau \phi_j(s')}\frac{\langle j(s')|\frac{dH(s')}{ds'}|0(s')\rangle}
{\Delta_{j0}(s')^2}\right]_0^s\right. \notag\\
&\qquad \left.-\frac{1}{\tau}\int_0^s ds'\, e^{\tau \phi_j(s')}
\frac{d}{ds'}\left(\frac{\langle j(s')|\frac{dH(s')}{ds'}|0(s')\rangle}
{\Delta_{j0}(s')^2}\right) \right\}\\
&=\frac{c_0(0)}{\tau}\frac{\langle j(s)|\frac{dH(s)}{ds}|0(s)\rangle}{\Delta_{j0}(s)^2}
+\mathcal{O}(\tau^{-2})\\
&\equiv \frac{c_0(0) A_j(s)}{\tau}+\mathcal{O}(\tau^{-2}) .
\label{cjne0}
\end{align}
From this and Eq.~(\ref{integralequation}), we obtain
\begin{align}
c_0^{(1)}(s)&=c_0(0)-\frac{c_0(0)}{\tau}\sum_{k\ne 0} \int_0^s ds'\,
\frac{\left| \langle k(s')|\frac{dH(s')}{ds'}|0(s')\rangle \right|^2}
{\Delta_{k0}(s')^3}
+\mathcal{O}(\tau^{-2})\\
&\equiv c_0(0)\left(1-\frac{1}{\tau} \int_0^s ds'\, B(s')\right)
+\mathcal{O}(\tau^{-2}) .
\label{c0}
\end{align}
Equations (\ref{cjne0}) and (\ref{c0}) represent the adiabatic approximation for the imaginary-time Schr\"odinger equation.

Equations (\ref{cjne0}) and (\ref{c0})  suggest that
the adiabatic condition for the imaginary-time Schr\"odinger equation is 
\begin{align}
\left|\frac{A_j(s)}{\tau}\right| \ll 1,\qquad\left|\frac{\int_0^s ds'B(s')}{\tau}\right| \ll 1,
\end{align}
the former of which coincides with the conventional adiabatic condition of the
real-time Schr\"odinger equation \cite{Messiah}.
We must be careful, however, that the norm of the wave function is not conserved,
and hence $|c_j(s)|^2$ does not directly represent the probability.
We shall come back to this point later.

\section{Master Equation Expressed as the Imaginary-Time Schr\"odinger Equation}
\label{sec:CtoQ}

Nonequilibrium dynamics of the Ising model following the master equation
can be rewritten as the imaginary-time Schr\"odinger equation as
described in  Refs.~\citen{Somma} and \citen{SA-QA}.
The master equation is
\begin{equation}
\frac{1}{\tau}\frac{dP_{\sigma}(s)}{ds}=\sum_{\sigma'}W_{\sigma\sigma'}(s)P_{\sigma'}(s),
\label{masterequation}
\end{equation}
where we have scaled the time as $s=t/\tau$ as before, $ \sigma $ is a set of $ N $ Ising spins $ \sigma=\{\sigma_1,\sigma_2,\dots,\sigma_N\} $, and $ P_{\sigma}(s) $ is the probability
that the system is in state $ \sigma $ at scaled time $ s $.
We have the Ising model with the Hamiltonian $H_0(\sigma)$ in mind, which
is reflected in the transition matrix $W_{\sigma\sigma'}(s)$ implicitly.
Note that the transition matrix $W_{\sigma\sigma'}(s)$ may be time-dependent
through the time dependence of the temperature $T(s)$ or its inverse $\beta (s)$.

Suppose that the transition matrix follows the detailed balance condition
\begin{equation}
W_{\sigma\sigma'}(s)P_{\sigma'}^{(0)}(s)=W_{\sigma'\sigma}(s)P_{\sigma}^{(0)}(s)
\quad \left(P_{\sigma}^{(0)}(s)=\frac{e^{-\beta (s)H_0(\sigma)}}{Z}\right).
\end{equation}
The right eigenvalues of the transition matrix are denoted as
$\lambda_0(=0)>\lambda_1>\lambda_2>\cdots$.
The leading eigenvalue/eigenvector corresponds to thermal equilibrium, which
does not change with time as suggested by $\lambda_0=0$.

The following `similarity transformation' is the key to mapping the classical
nonequilibrium dynamics to quantum mechanics \cite{Henley,Castelnovo,Somma,SA-QA},
\begin{align}
H_{\mathrm{ SA}}(s)&\equiv -e^{\frac{1}{2}\beta(s) H_0}W(s)e^{-\frac{1}{2}\beta (s)H_0}
\label{HSA}\\
|\psi (s)\rangle & \equiv e^{\frac{1}{2}\beta (s)H_0}\sum_{\sigma}P_{\sigma}|\sigma\rangle
\label{phi},
\end{align}
where $W(s)$ is a $ 2^N\times 2^N $ matrix with elements $W_{\sigma\sigma'}(s)$,
and $H_{\mathrm{ SA}}(s)$ is also a matrix.
Note that $|\psi (s)\rangle$ is not normalized.
It is easy to see that this $H_{\mathrm{ SA}}(s)$ is Hermitian,
and can therefore be regarded as a quantum-mechanical Hamiltonian.
Two matrices, $W(s)$ and $H_{\mathrm{ SA}}(s)$, share the spectrum and eigenstates,
up to a trivial factor or sign, 
\begin{align}
W(s)|\lambda_n(s)\rangle &=\lambda_n(s)|\lambda_n(s)\rangle \\
H_{\mathrm{ SA}}(s)|\phi^{(n)}(s)\rangle &=E_{\mathrm{ SA}}^{(n)}(s)|\phi^{(n)}(s)\rangle
=-\lambda_n(s)|\phi^{(n)}(s)\rangle\\
|\phi^{(n)}(s)\rangle&=e^{\frac{1}{2}\beta(s)H_0} |\lambda_n(s)\rangle,
\end{align}
as can be verified from Eqs. (\ref{HSA}) and (\ref{phi}).
The vectors $|\psi (s)\rangle$ and $|\phi^{(n)}(s)\rangle$ are not normalized.
The normalized eigenvector of $H_{\mathrm{ SA}}(s)$ will be denoted as $|n_{\mathrm{ SA}}(s)\rangle$.
In particular, the normalized ground state is
\begin{equation}
|0_{\mathrm{ SA}}(s)\rangle =\frac{e^{-\frac{1}{2}\beta(s) H_0}}{\sqrt{Z}}\sum_{\sigma}|\sigma \rangle,
\end{equation}
which corresponds to thermal equilibrium having $\lambda_0(s)=0$
and consequently $E_{\mathrm{ SA}}^{(0)}(s)=0$.
The expectation value of an arbitrary matrix diagonal in the $ \sigma $-basis by the ground state of $H_{\mathrm{ SA}}(s)$
is equal to the expectation value by the Boltzmann distribution.
This suggests that thermal fluctuations are mapped to quantum fluctuations of the ground state.

From the master equation (\ref{masterequation}),  $|\psi (s)\rangle$ can be verified
to satisfy the following differential equation,
\begin{equation}
-\frac{d}{ds}|\psi (s)\rangle =\tau\left(H_{\mathrm{ SA}}(s)-\frac{1}{2\tau}\dot{\beta}(s)H_0
\right) |\psi (s)\rangle,
\end{equation}
where $\dot{\beta}$ is for $d\beta /ds$.
This is  a type of imaginary-time Schr\"odinger equation with the effective
Hamiltonian
\begin{equation}
H_{\mathrm{ tot}}(s)\equiv H_{\mathrm{ SA}}(s)-\frac{\dot{\beta}(s)}{2\tau}H_0.
\label{totalhamiltonian}
\end{equation}
The normalized instantaneous eigenstate of the effective Hamiltonian will be written as
\begin{equation}
H_{\mathrm{ tot}}(s)|j_{\mathrm{ tot}}(s)\rangle =E_{\mathrm{ tot}}(s)|j_{\mathrm{ tot}}(s)\rangle .
\end{equation}

\section{Probability of Reaching the Ground State}
\label{sec:prob}

Let us write the spin configuration of the ground state of $H_0$ as $|\sigma_{\mathrm{ G}}\rangle$.
The probability that the system reaches the ground state at time $s$ is
\begin{align}
P_{\sigma_{\mathrm{ G}}}(s)&=\langle \sigma_{\mathrm{ G}}|\sum_{\sigma}P_{\sigma}(s)|\sigma\rangle\\
&=\langle \sigma_{\mathrm{ G}}|e^{-\frac{1}{2}\beta (s)H_0}|\psi (s)\rangle\\
&=e^{-\frac{1}{2}\beta (s) E_{\mathrm{ G}}} \langle \sigma_{\mathrm{ G}}|\psi (s)\rangle .
\end{align}
This expression can be decomposed as
\begin{align}
P_{\sigma_{\mathrm{ G}}}(s)
&=e^{-\frac{1}{2}\beta (s)E_{\mathrm{ G}}}\sum_{j,k} \langle \sigma_{\mathrm{ G}}|k_{\mathrm{ SA}}(s)\rangle
\langle k_{\mathrm{ SA}}(s)|j_{\mathrm{ tot}}(s)\rangle \langle j_{\mathrm{ tot}}(s)|\psi (s)\rangle .
\end{align}
Now, we assume that the temperature is controlled such that it reaches $T=0~(\beta\to\infty)$
at $s=1$ and that the ground-state energy of $H_0$ is also zero, $E_{\mathrm{ G}}=0$.
Then, the instantaneous eigenstate of $H_{\mathrm{ SA}}(s)$ at $s=1$ is the ground state, so
$\langle \sigma_{\mathrm{ G}}|k_{\mathrm{ SA}}(1)\rangle =\delta_{k,0}$.
We therefore have
\begin{equation}
P_{\sigma_{\mathrm{ G}}}(1)=\sum_{j}
\langle 0_{\mathrm{ SA}}(1)|j_{\mathrm{ tot}}(1)\rangle \langle j_{\mathrm{ tot}}(1)|\psi (1)\rangle .
\label{Psg}
\end{equation}
According to the definition  (\ref{totalhamiltonian}) and perturbation theory, the
instantaneous eigenstate
of the total Hamiltonian is related in the large-$\tau$ limit to $H_{\mathrm{ SA}}$ as
\begin{equation}
|j_{\mathrm{ tot}}\rangle =|j_{\mathrm{ SA}}\rangle -\frac{\dot{\beta}}{2\tau}
\sum_{l\ne j}|l_{\mathrm{ SA}}\rangle \frac{\langle l_{\mathrm{ SA}}|H_0|j_{\mathrm{ SA}}\rangle}
{E_{\mathrm{ SA}}^{(j)}(s)-E_{\mathrm{ SA}}^{(l)}(s)}+\mathcal{O}(\tau^{-2}) .
\end{equation}
We thus have
\begin{equation}
\langle 0_{\mathrm{ SA}}(s)|0_{\mathrm{ tot}}(s)\rangle =1+\mathcal{O}(\tau^{-2}),
\label{projection00}
\end{equation}
and
\begin{equation}
\langle 0_{\mathrm{ SA}}(s)|j_{\mathrm{ tot}}(s)\rangle =-\frac{\dot{\beta}}{2\tau}
\frac{ \langle 0_{\mathrm{ SA}}|H_0|j_{\mathrm{ SA}}\rangle}{E_{\mathrm{ SA}}^{(j)}(s)-E_{\mathrm{ SA}}^{(0)}(s)}
+\mathcal{O}(\tau^{-2})\quad (j\ne 0).
\end{equation}
Then, from Eq.~(\ref{Psg}), 
\begin{align}
P_{\sigma_{\mathrm{ G}}}(1)&=\langle 0_{\mathrm{ tot}}(1)|\psi (1)\rangle-\frac{\dot{\beta}}{2\tau}\sum_{j\ne 0} \frac{\langle 0_{\mathrm{ SA}}|H_0|j_{\mathrm{ SA}}\rangle}
{E_{\mathrm{ SA}}^{(j)}(s)-E_{\mathrm{ SA}}^{(0)}(s)}\langle j_{\mathrm{ tot}}(1)|\psi (1)\rangle
+\mathcal{O}(\tau^{-2}).
\end{align}
The asymptotic expansions of Eqs. (\ref{cjne0}) and (\ref{c0}) developed in Sect.~\ref{sec:IS}
for the imaginary-time Schr\"odinger equation tell us that
\begin{align}
\langle 0_{\mathrm{ tot}}(s)|\psi(s)\rangle &=c_{0}^{\mathrm{ tot}}(0)\left(
1-\frac{1}{\tau} \int_0^s
B_{\mathrm{ tot}}(s')ds'\right) +\mathcal{O}(\tau^{-2}) \\
\langle j_{\mathrm{ tot}}(s)|\psi (s)\rangle &=\frac{c_{0}^{\mathrm{ tot}}(1)\,
A_{\mathrm{ tot}}^{(j)}}{\tau}+\mathcal{O}(\tau^{-2}),
\end{align}
from which we have
\begin{equation}
P_{\sigma_{\mathrm{ G}}}(1)=c_{0}^{\mathrm{ tot}}(0)\left(
1-\frac{1}{\tau}\int_0^1 B_{\mathrm{ tot}}(s)ds\right)+\mathcal{O}(\tau^{-2}),
\label{psg01}
\end{equation}
where
\begin{equation}
B_{\mathrm{ tot}}(s)=\sum_{j\ne 0} \frac{\left|\langle j_{\mathrm{ tot}}(s)|
\frac{dH_{\mathrm{ tot}}(s)}{ds}|0_{\mathrm{ tot}}(s)\rangle \right|^2}{(E_{\mathrm{ tot}}^{(j)}(s)-E_{\mathrm{ tot}}^{(0)}(s))^3}.
\end{equation}
Since the difference between $H_{\mathrm{ tot}}$ and $H_{\mathrm{ SA}}$ is of $\mathcal{O}(\tau^{-1})$,
we finally obtain
\begin{equation}
P_{\sigma_{\mathrm{ G}}}(1)
=c_{0}^{\mathrm{ SA}}(0)\left(
1-\frac{1}{\tau}\int_0^1 B_{\mathrm{ SA}}(s)ds\right)+\mathcal{O}(\tau^{-2}),
\label{psg00}
\end{equation}
where
\begin{equation}
B_{\mathrm{ SA}}(s)=\sum_{j\ne 0} \frac{\left|\langle j_{\mathrm{ SA}}(s)|
\frac{dH_{\mathrm{ SA}}(s)}{ds}|0_{\mathrm{ SA}}(s)\rangle \right|^2}{E_{\mathrm{ SA}}^{(j)}(s)^3},
\end{equation}
and we have replaced $c_{0}^{\mathrm{ tot}}(1)$ by $c_{0}^{\mathrm{ SA}}(1)$
because the difference of these coefficients is of $\mathcal{O}(\tau^{-2})$
according to Eq.~(\ref{projection00}).
We hereafter assume $c_{0}^{\mathrm{ SA}}(0)=1$,
which indicates that the initial state was the ground state of $H_{\mathrm{ SA}}(0)$,
\textit{i.e.},  the thermal equilibrium state at the inverse temperature $ \beta(0) $  .

We have also taken into account the fact that the ground-state energy of $H_{\mathrm{ SA}}$ is
$E_{\mathrm{ SA}}^{(0)}(s)=0$,
\begin{equation}
H_{\mathrm{ SA}}(s)\left( e^{-\frac{1}{2}\beta(s)H_0} \sum_{\sigma} |\sigma \rangle \right)=0.
\end{equation}
In order to simplify the expression for $B_{\mathrm{ SA}}(s)$, we take the derivative of the
above equation with respect to $s$,
\begin{align}
&\frac{dH_{\mathrm{ SA}}(s)}{ds}\left( e^{-\frac{1}{2}\beta (s)H_0}
\sum_{\sigma}|\sigma \rangle \right)=H_{\mathrm{ SA}}(s)\left( \frac{1}{2}\dot{\beta} (s)H_0 e^{-\frac{1}{2}\beta (s)
H_0} \sum_{\sigma} |\sigma \rangle \right).
\end{align}
The projection of this equation to $|j_{\mathrm{ SA}}(s)\rangle$ gives
\begin{equation}
\langle j_{\mathrm{ SA}}(s)|\frac{dH_{\mathrm{ SA}}(s)}{ds}|0_{\mathrm{ SA}}(s)\rangle
=\frac{E_{\mathrm{ SA}}^{(j)} \dot{\beta}(s)}{2} \langle j_{\mathrm{ SA}}(s)
|H_0 |0_{\mathrm{ SA}}(s)\rangle,
\end{equation}
from which we have the simplified expression
\begin{equation}
B_{\mathrm{ SA}}(s)=\frac{\dot{\beta}^2}{4}\sum_{j \ne 0}
\frac{|\langle j_{\mathrm{ SA}}(s)|H_0 |0_{\mathrm{ SA}}(s)\rangle |^2}
{E_{\mathrm{ SA}}^{(j)}(s)}.
\label{BSA_simple}
\end{equation}

\section{Convergence Condition of Simulated Annealing}
\label{sec:cc}

We are now ready to analyze the problems of the analysis in Somma~\textit{et al.}~\cite{Somma}.
They used the classical-to-quantum mapping described in Sect.~\ref{sec:CtoQ} to rewrite
classical nonequilibrium dynamics as quantum mechanics. 
Then they applied the conventional adiabatic condition for the real-time
Schr\"odinger equation to derive a differential equation for the temperature
variable of the original classical system.
By solving this differential equation, they `rederived' the Geman-Geman~\cite{Geman}
condition
\begin{equation}
T(t)\approx \frac{pN}{\log t},
\end{equation}
for the original classical dynamics of the
Ising model to reach the ground state with probability
close to unity in the limit of the long time scale, $t \gg \mathcal{O}(1/\Delta)$,
where $\Delta$ is $|\lambda_1(s)|=E_{\mathrm{ SA}}^{(1)}(s)$ in our notation.
The quantity in the numerator $p$ is an $\mathcal{O}(1)$ constant.

There are two points of incompleteness in their argument.
First, we have to use the imaginary-time Schr\"odinger equation to
analyze classical dynamics, not the real-time Schr\"odinger equation.
The adiabatic conditions of these two cases have subtle differences
as discussed in detail in Sect.~\ref{sec:IS}.
The second problem is that they did not use
the exact expression for the mapped Hamiltonian $H_{\mathrm{SA}}(s)$ defined in Eq.~(\ref{HSA})
but replaced it by a simpler form with the coefficient of the transverse-field
term being constant,
\begin{equation}
H_{q}'=H-\chi \sum_j \sigma_x^j,
\end{equation}
in their notation, where $\chi =e^{-p\beta}$.

Let us discuss the second point first since it is not a very serious one.
According to Eq.~(\ref{HSA}),
the mapped Hamiltonian $H_{\mathrm{ SA}}$ is a generalized transverse-field
Ising model where the coefficient of the transverse field generally
depends on the spin configuration.  For example, the simplest
case of the one-dimensional Ising model is mapped to \cite{SA-QA} 
\begin{align}
H_{\mathrm{ SA}}&=\frac{N}{2}-\frac{1}{2}\tanh 2\beta J 
\sum_{j=1}^N \sigma_j^z \sigma_{j+1}^z-\frac{1}{2\cosh 2\beta J }\sum_{j=1}^N
\big(\cosh^2\beta J  -\sinh^2 \beta J  \, \sigma_{j-1}^z \sigma_{j+1}^z\big)\sigma_j^x.
\label{hhat-hb}
\end{align}
$\sigma^z$-dependence exists in the coefficient of $\sigma^x$.
Nevertheless, for the purpose of evaluation of the smallest energy gap,
it is allowed to replace the coefficients by their smallest
values, which depend on $\beta$ exponentially as $e^{-p\beta}(=\chi)$.
The reason for this is that the evaluation of the smallest energy gap using
the Hopf theorem~\cite{Hopf}, as discussed in Somma~\textit{et al.}~\cite{Somma} and as described in
detail in Lemma 3.3 of Morita and Nishimori~\cite{Morita},
uses only the smallest values of the off-diagonal elements.
Thus, the resulting general lower bound of the energy gap
\begin{equation}
\Delta(s)=E_{\mathrm{ SA}}^{(1)}(s)\ge a\sqrt{N}e^{-2(p\beta (s)+c)N},
\label{gapbound}
\end{equation}
where $ a $ and $ c $ are $ N $-independent positive constants,
can be used in the present context.

The first point regarding the difference between imaginary-time and real-time
Schr\"odinger dynamics must be taken more seriously, for which reason we
have developed a theory of the previous sections.
If we are allowed to ignore higher-order terms than the first order in $\tau^{-1}$,
which itself needs verification rigorously speaking, the condition
that the ground-state probability is sufficiently close to unity is,
according to Eqs.~(\ref{psg00}) and (\ref{BSA_simple}),
\begin{align}
\left| \frac{1}{\tau}\int_0^1 B_{\mathrm{ SA}}(s)ds\right|\ll 1\\
B_{\mathrm{ SA}}(s)=\frac{\dot{\beta}^2}{4}\sum_{j \ne 0}
\frac{|\langle j_{\mathrm{ SA}}(s)|H_0 |0_{\mathrm{ SA}}(s)\rangle |^2}
{E_{\mathrm{ SA}}^{(j)}(s)}.
\end{align}
To satisfy this condition, the largest term in the above sum
(with $j=1$) must be very small. 
If we replace the denominator of the expression for $B_{\mathrm{ SA}}(s)$
by its smallest value in Eq.~(\ref{gapbound}) and the matrix element
in the numerator by its upper bound, a constant times the system size $pN$,
we obtain the following condition:
\begin{equation}
\frac{4e^{2cN}p^2 N^2}{a\sqrt{N}} \int_0^{\tau} (\dot{\beta})^2 e^{2\beta pN}dt=\delta \ll 1,
\end{equation}
where we have restored the original time scale $t=s\tau$.
The dot over $\beta$ now denotes the derivative with respect to $t$.
We next take the limit of the infinite time scale, $\tau\to\infty$, which
is the situation for which the Geman-Geman condition was originally derived.
Then, only the upper bound of the above integral relation is changed to
infinity provided that $\beta$ is a function of $t$ only,
\textit{i.e.}, without $\tau$-dependence.
For the resulting condition
\begin{equation}
\frac{4e^{2cN}p^2 N^2}{a\sqrt{N}} \int_0^{\infty} (\dot{\beta})^2 e^{2\beta pN}dt=\delta \ll 1
\end{equation}
to hold,  the integrand should approach zero sufficiently quickly in
the large-$t$ limit.
More explicitly, $\beta (t)$ is expected to asymptotically
satisfy the differential equation
\begin{equation}
\frac{4e^{2cN}p^2 N^2}{a\sqrt{N}}(\dot{\beta})^2
 e^{2\beta pN}=b^2 t^{-1-\epsilon}~(\epsilon >0), \label{11}
\end{equation}
with sufficiently small $b$.
By rewriting the above as 
\begin{equation}
\frac{2e^{cN}pN}{\sqrt{a\sqrt{N}}}\frac{d\beta}{dt}e^{\beta pN}=b t^{-(1+\epsilon)/2},
\end{equation}
we solve it for $\beta (t)$ as
\begin{equation}
\frac{2e^{cN}}{\sqrt{a\sqrt{N}}}e^{\beta pN}=\frac{2b}{1-\epsilon} t^{(1-\epsilon)/2}+c',
\end{equation}
or
\begin{align}
\beta pN&=-cN +\frac{1}{2}\log (a\sqrt{N})-\log 2
+\log \left( \frac{2b}{1-\epsilon}t^{(1-\epsilon)/2}+c'\right).
\label{detailed-eq}
\end{align}
If we keep only the leading-order term for large $t$,
\begin{equation}
\beta(t)\approx\frac{1-\epsilon}{2pN}\log t.
\end{equation}
This agrees with Somma~\textit{et al.} except for a small correction $\epsilon (>0)$.
Notice that their $pN$ is our $2pN$.

\section{Conclusion}
\label{sec:Con}

We have established adiabatic-theorem-like relations for the imaginary-time
Schr\"odinger dynamics.
This was done before in Ref.~\citen{Morita} for norm-conserved dynamics,
which is not necessarily suitable for the analysis of the master equation
of classical Markovian dynamics.
De Grandi~\textit{et al.}~\cite{Grandi}  also discussed this problem.
We developed their calculations further to obtain a more compact expression, as seen in Eqs.~(\ref{cjne0}) and (\ref{c0}).
The result was applied to studying the validity of the analysis in Ref.~\citen{Somma},
which rederived the convergence condition of simulated annealing to the target ground state.
We have found that the conclusion of Ref.~\citen{Somma} is correct,
but the process to reach it needs more careful analyses as developed here.
Our theoretical framework may also be used to shed new light on
the analysis of finite-temperature slow dynamics of classical Ising models,
\textit{e.g.}, spin glasses.

\end{document}